\documentclass[conference]{IEEEtran}
\IEEEoverridecommandlockouts
\usepackage{cite}
\usepackage{amsmath,amssymb,amsfonts}
\usepackage{algorithmic}
\usepackage{graphicx}
\usepackage{textcomp}
\usepackage{pifont}
\usepackage{xcolor}

\usepackage{xspace}
\usepackage{hyperref}
\usepackage{booktabs}
\usepackage{subcaption}
\newcommand{\stitle}[1]{\vspace{0em}\textit{\underline{#1.}}}
\newcommand{\gv}{GoVector\xspace}
\newcommand{\eat}[1]{}
\def\BibTeX{{\rm B\kern-.05em{\sc i\kern-.025em b}\kern-.08em
    T\kern-.1667em\lower.7ex\hbox{E}\kern-.125emX}}
\begin{document}

\title{GoVector: An I/O-Efficient Caching Strategy for High-Dimensional Vector Nearest Neighbor Search\\
\thanks{The first two authors contributed equally to this paper.}

\thanks{This paper is the English version of our Chinese paper accepted for publication in \textit{Journal of Software}, Vol.~37, No.~3, 2026.}
}

\author{
    Yijie Zhou, Shengyuan Lin, Shufeng Gong, Song Yu, Shuhao Fan, Yanfeng Zhang, Ge Yu \\
    \textit{Northeastern University} \\
    \{zhouyijie, linshengyuan, yusong, 3074987576\}@stumail.neu.edu.cn \\
    \{gongsf, zhangyf, yuge\}@mail.neu.edu.cn
}

\maketitle

\begin{abstract}
Graph-based high-dimensional vector indices have become a mainstream solution for large-scale approximate nearest neighbor search (ANNS). However, their substantial memory footprint often requires storage on secondary devices, where frequent on-demand loading of graph and vector data leads to I/O becoming the dominant bottleneck, accounting for over 90\% of query latency. Existing static caching strategies mitigate this issue only in the initial navigation phase by preloading entry points and multi-hop neighbors, but they fail in the second phase where query-dependent nodes must be dynamically accessed to achieve high recall.
We propose GoVector, an I/O-efficient caching strategy tailored for disk-based graph indices. GoVector combines (1) a static cache that stores entry points and frequently accessed neighbors, and (2) a dynamic cache that adaptively captures nodes with high spatial locality during the second search phase. To further align storage layout with similarity-driven search patterns, GoVector reorders nodes on disk so that similar vectors are colocated on the same or adjacent pages, thereby improving locality and reducing I/O overhead.
Extensive experiments on multiple public datasets show that GoVector achieves substantial performance improvements. At 90\% recall, it reduces I/O operations by 46\% on average, increases query throughput by 1.73$\times$, and lowers query latency by 42\% compared to state-of-the-art disk-based graph indexing systems.
\end{abstract}


\section{Introduction}\label{sec:intro}

The rapid advancement of generative AI, exemplified by Large Language Models (LLMs), has made vector-based semantic retrieval a core component in natural language processing \cite{asai2023retrieval, li2023skillgpt}, information retrieval\cite{grbovic2018real,huang2020embedding}, and recommendation systems\cite{okura2017embedding,covington2016deep}. In Retrieval-Augmented Generation (RAG), vector retrieval plays a crucial role in improving generation quality and response efficiency\cite{gao2023retrieval}. To deal with the complexity of high-dimensional vector search\cite{gao2023high}, Approximate Nearest Neighbor Search (ANNS) has been widely adopted. Among existing approaches, graph-based indices such as HNSW\cite{malkov2018efficient}, NSG\cite{fu2017fast}, and DiskANN\cite{jayaram2019diskann} have attracted widespread attention from both academia and industry for their low latency, high accuracy, and high throughput\cite{10.5555/3540261.3540659,sun2014srs,bentley1975multidimensional,arora2018hd}.

\begin{figure}[t]
    \centering
    \includegraphics[width=.9\linewidth]{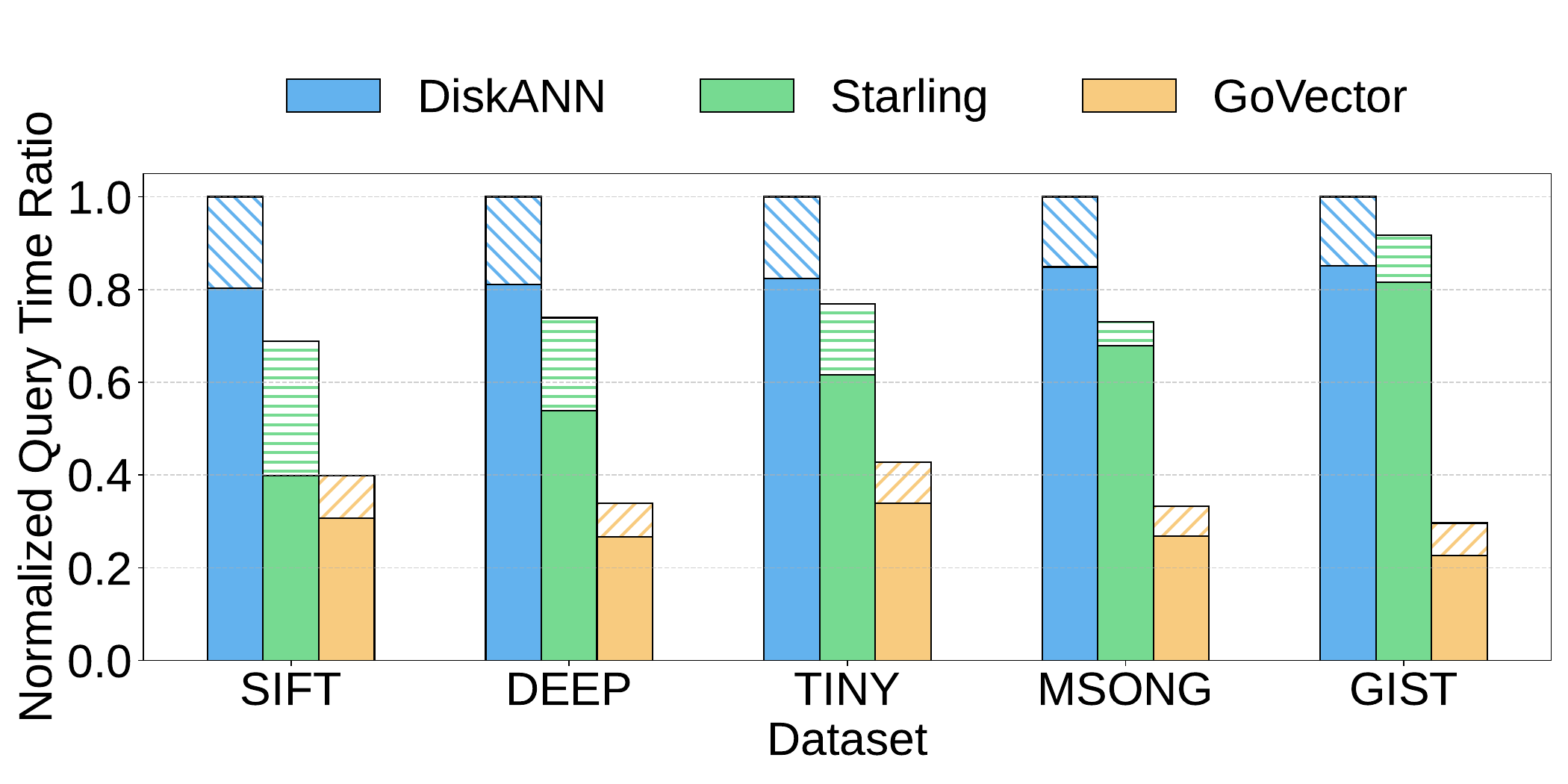} 
    \caption{Comparison of normalized I/O-CPU query time ratio across ANNS systems.}\label{fig:cpu-io-ratio}
\end{figure}

However, graph-based indices incur substantial storage costs due to explicit maintenance of large-scale adjacency lists. As vector datasets scale, keeping the entire index in memory becomes prohibitively expensive, prompting the adoption of disk-based ANNS solutions that store part or all of the index graph on secondary storage\cite{jayaram2019diskann,wang2024starling}. For example, DiskANN has been integrated into high-performance vector database systems such as Pinecone\cite{pinecone2025} and Milvus\cite{wang2021milvus} to support efficient retrieval at the scale of hundreds of billions of vectors. These methods typically start from an entry node, iteratively expanding by loading the corresponding disk page to obtain its vector and adjacency information, and selecting the next expansion node based on neighbor-query distances.

Despite their scalability, disk-based graph indices suffer from severe I/O bottlenecks. We conducted a detailed analysis of the query performance of two state-of-the-art disk-based ANNS systems, DiskANN and Starling, on five real-world datasets\cite{jegou2009searching,simhadri2022results,yang2024fast}. Fig. \ref{fig:cpu-io-ratio} presents the proportion of query time spent on I/O operations and CPU computation at a 90\% recall rate. The results show that I/O operations dominate query latency, accounting for an average of 83\% in DiskANN and 79\% in Starling, indicating that disk access has become the primary bottleneck limiting overall system performance.

The causes of this bottleneck can be attributed to two main factors. (1) The ANNS query path is highly query-dependent: its search process relies on the position of the query vector in the vector space, making it difficult to accurately predict the nodes to be accessed before the query begins. Consequently, mainstream systems generally adopt a static caching strategy, in which the entry node and several of its multi-hop neighbors are preloaded into memory in the hope that they will be directly hit during the search, thereby reducing disk reads\cite{jayaram2019diskann}. However, this strategy lacks awareness of the actual query path, leaving much of the cached data unused and resulting in a low overall hit rate. For example, in DiskANN, once the search enters the neighborhood region of the query vector, the cache hit rate during the top-k similarity search phase is only 4\%-9\%. (2) Existing systems typically optimize the disk layout of index graphs based on their topological structure to reduce I/O overhead. For instance, Starling\cite{wang2024starling}, one of the most advanced disk-based index graph locality optimization schemes, reduces random I/O operations by grouping nodes and their topological neighbors into the same disk page. However, this approach overlooks the unique nature of graph-based queries: the selection of expansion nodes during the query process does not strictly follow breadth-first or depth-first traversal, but is instead guided by their distances to the query vector. In summary, current disk-based graph index methods face two major challenges in reducing I/O overhead:

\begin{itemize}
    \item \textbf{Low cache hit rate.} Since the query path is highly query-dependent and dynamic, existing static caching strategies cannot capture the actual query path, making it difficult to achieve cache hits on the vertices along the path. It is necessary to design a flexible and query-adaptive cache mechanism that can selectively cache key vertices that may be accessed during the search process based on the characteristics of the query vector. Such a design can improve cache hit rates and reduce unnecessary disk access. 
    
    \item \textbf{Insufficient data utilization per I/O.} Topology-based index layouts fail to fully capture the access patterns of actual query paths, resulting in only a small portion of data being effectively utilized in each round of disk loading. To address this, the physical organization of the index should be redesigned by taking into account the search characteristics of ANNS, improving the locality of index graph access so that each I/O operation can load more query-relevant vectors and adjacency information, improving overall retrieval efficiency.
\end{itemize}

To address these challenges, we propose \textbf{\gv}, an I/O-efficient caching strategy for disk-based graph indices in high-dimensional vector search. \gv improves query efficiency through query-aware hybrid caching and a vector-similarity-based index reordering. The main contributions of this paper are as follows:

\begin{itemize}
    \item We propose a static-dynamic hybrid caching strategy. The static cache preloads the entry node and several of its neighbors to quickly guide the search toward the neighborhood of the query vector, while the dynamic cache adaptively retains candidate nodes and their similar neighbors encountered along the query path, thereby improving cache hit rates during the similarity search phase.
    
    \item We design a vector-similarity-based index layout optimization that stores highly similar vectors in the same or adjacent disk pages. This increases the effective data per I/O, reduces the number of disk accesses, and improves overall disk performance.

    \item We conduct a systematic evaluation on multiple public datasets. Compared to existing approaches, \gv reduces I/O operations by an average of 46\% (up to 57\%), increases query throughput by 1.73$\times$ (up to 2.25$\times$), and lowers query latency by 42\% (up to 55\%).
\end{itemize}

The remainder of this paper is organized as follows. Section \ref{sec:pre} introduces the preliminaries; Section \ref{sec:work} reviews background and related work; Section \ref{sec:overview} presents an overview of \gv and its system architecture; Sections \ref{sec:hybrid} and \ref{sec:reorder} detail the core designs, including the hybrid caching mechanism and index reordering; Section \ref{sec:expr} reports the experimental setup and evaluation results; and Section \ref{sec:con} concludes the paper.
\section{Preliminaries}\label{sec:pre}

\subsection{Approximate Nearest Neighbor Search}

Approximate Nearest Neighbor Search (ANNS) aims to quickly identify the top-$k$ vectors most similar to a given query vector within a high-dimensional vector collection\cite{fu2017fast,fu2021high,jegou2022faiss,wang2024starling}. Given a dataset $V = \{v_1, v_2, \ldots, v_n\} \subset \mathbb{R}^d$, containing $n$ high-dimensional vectors and a query vector $q \in \mathbb{R}^d$, the goal of ANNS is to find the $k$ vectors in $V$ closest to $q$. Formally, the result set can be defined as: $\text{Top}_k(q) = \arg\min_{\substack{R \subseteq V, |R| = k}} \sum_{v \in R} d(q, v)$, where $d(\cdot,\cdot)$ denotes the distance function in the vector space, commonly chosen as Euclidean distance or cosine similarity, and $R$ represents the candidate top-$k$ result set.

Recall is one of the key metrics for evaluating the quality of vector search, measuring how many of the true nearest neighbors are included in the returned results. Let $GT_k(q)$ denote the ground-truth top-k nearest neighbor set of $q$, and $ANN_k(q)$ the approximate result set returned by the system. Then the recall at top-k is defined as: $\text{Recall@}k(q) = {|ANN_k(q) \cap GT_k(q)|}/{k}$. The recall value ranges from $[0, 1]$, with values closer to 1 indicating more accurate results. The objective of ANNS is to minimize query latency and resource consumption (e.g., memory and I/O operations) while maintaining high recall, thereby enabling scalable and efficient vector retrieval in large-scale high-dimensional datasets.

\subsection{Graph-Based ANNS}

\begin{figure}[t]
    \centering
    \begin{subfigure}{.32\linewidth}
        \centering
        \includegraphics[width=\linewidth]{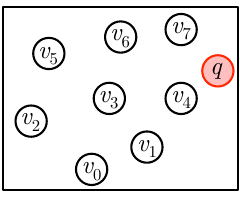}
        \caption{Original vectors}\label{fig:pre-eg1}
    \end{subfigure}
    \begin{subfigure}{.32\linewidth}
        \centering
        \includegraphics[width=\linewidth]{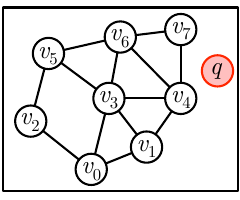}
        \caption{Graph index}\label{fig:pre-eg2}
    \end{subfigure}
    \begin{subfigure}{.32\linewidth}
        \centering
        \includegraphics[width=\linewidth]{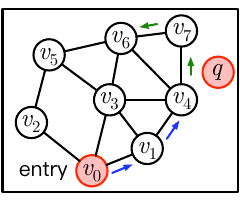}
        \caption{Search process}\label{fig:pre-eg3}
    \end{subfigure}
    \caption{Illustration of ANNS based on graph index.}
    \label{fig:pre-eg}
\end{figure}

In ANNS, the graph index is an indexing structure that strikes a balance between search accuracy and efficiency\cite{wang2024starling}. Its core idea is to construct a graph from the original high-dimensional vectors, where similar vectors are connected by edges. The search process then approaches the query vector’s neighbors step by step by traversing along these edges. Formally, given a vector set $\mathcal{V}=\{v_1,v_2,\ldots,v_n\}\subset\mathbb{R}^d$, where each $v_i$ denotes the $i$-th vector with coordinate $x_{v_i}\in\mathbb{R}^d$, a neighborhood graph $G=(V,E)$ is constructed. Each node corresponds to a vector, while the edge set $E$ connects each node with a subset of its neighbors, forming the index graph structure. As illustrated in Fig. \ref{fig:pre-eg}, panel \ref{fig:pre-eg2} shows the index graph constructed from the dataset in panel \ref{fig:pre-eg1}.

In graph-based ANNS, given a query vector $q$, the algorithm first selects an entry point $v_{\text{start}}$ either randomly or via a heuristic strategy. Starting from this node, its neighbors are inserted into a candidate queue $L$ according to their distances to $q$. At each iteration, the algorithm expands the closest unvisited node $p^\ast$ from $L$ (retrieving its vector and neighbor information), until the top-$l$ nodes in $L$ have all been visited. Let the index graph be denoted as $G(P,E)$, where the node set $P$ represents all data points and the edge set $E$ defines their connectivity. For any node $p \in P$, let its vector be $x_p$, the query vector be $x_q$, and the distance function be Euclidean distance $dist(p,q)=\|x_p-x_q\|_2$. If the candidate set is $L\subseteq P$ and the visited set is $S\subseteq P$, the node expanded in each iteration is selected as: $p^\ast = \arg\min_{p \in L \setminus S} d(p,q)$. The candidate queue and visited set are updated as $L \gets L \cup N_{out}(p^\ast)$, $S \gets S \cup \{p^\ast\}$, where $N_{out}(p^\ast)$ denotes the neighbors of $p^\ast$. This process repeats until $|L \cap S|=l$. Finally, the top-$k$ ($k \le l$) closest nodes to $q$ in $L$ are returned as the result set.

This strategy is referred to as GreedySearch, and in practice it is often improved into BeamSearch to enhance parallel I/O efficiency. For example, in Fig. \ref{fig:pre-eg3}, the search starts from entry point $v_0$. Its three neighbors are ordered by distance to $q$ as $v_1$, $v_3$, and $v_2$, and are sequentially inserted into $L$. The closest node $v_1$ is then expanded. Since $v_3$ is already in $L$, only $v_1$’s other neighbor $v_4$ is added. At this point, among the candidates $\{v_2,v_3,v_4\}$, $v_4$ is closest to $q$ and is selected for the next expansion. The algorithm thus successfully locates $q$’s nearest neighbor $v_4$. To retrieve the top-$k$ nearest neighbors of $q$, the algorithm continues this procedure until the termination condition is met.

\subsection{Two-Phase Decomposition of the Search Process}\label{subsec:two=phase}

\begin{figure}[t]
    \centering
    \begin{subfigure}{.49\linewidth}
        \centering
        \includegraphics[width=\linewidth]{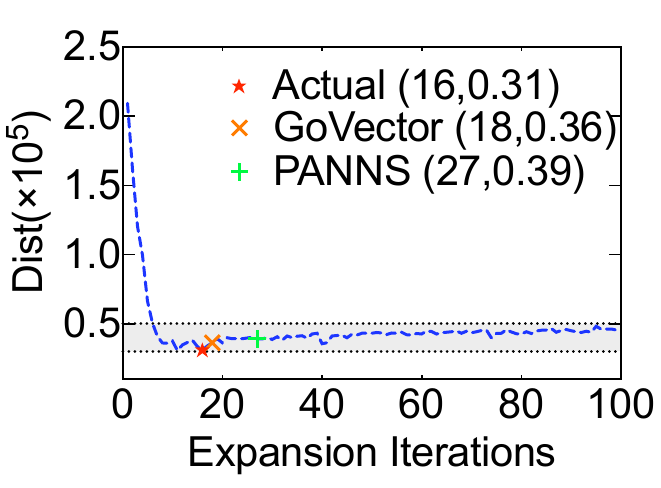}
        \caption{SIFT}\label{fig:pre-stage1}
    \end{subfigure}
    \hfill
    \begin{subfigure}{.49\linewidth}
        \centering
        \includegraphics[width=\linewidth]{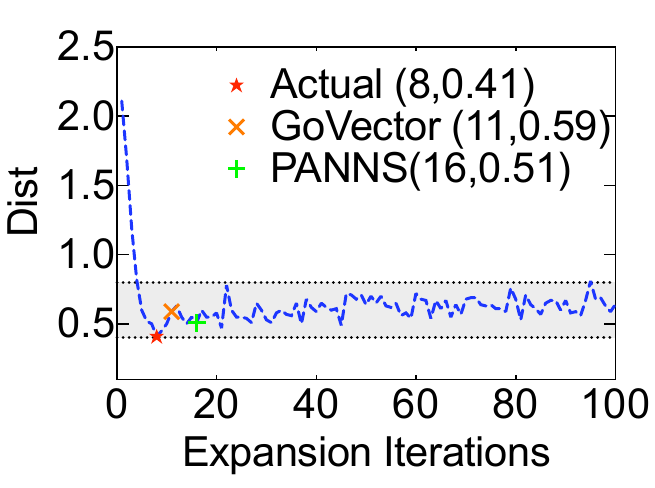}
        \caption{GIST}\label{fig:pre-stage2}
    \end{subfigure}
    \caption{Distance variation between expansion node $p^\ast$ and query vector $q$ during ANNS queries on different datasets ($k=10$).}
    \label{fig:pre-stage}
\end{figure}

The query process of ANNS exhibits a clear two-phase characteristic\cite{yin2025panns}. As shown in Fig. \ref{fig:pre-eg3}, after locating the nearest neighbor $v_4$ of query vector $q$, the search must continue to expand in order to explore the remaining candidate nodes and return the complete top-$k$ results. At this phase, the expansion path may gradually drift away from the query point $q$, causing distances to increase overall. However, since the search is still constrained to several hops within the local candidate region and expansions tend to prioritize closer nodes, the distance values usually fluctuate within a relatively narrow range. Fig. \ref{fig:pre-stage} illustrates this phenomenon on the SIFT and GIST datasets (each with one million vectors), showing how the distance between query vector $q$ and the expansion node $p^\ast$ evolves with each expansion step. Red dots mark the iteration where the nearest neighbor of the query is first reached, along with the corresponding minimum distance. It can be observed that before reaching this point, the distance decreases rapidly, while afterwards the distance variations stabilize into a narrower fluctuation range. Based on this observation, we characterize ANNS queries using a two-phase concept.

The first phase is a rapid convergence phase, where the search quickly approaches the query vector. Most graph- or tree-based ANNS algorithms (e.g., HNSW, NSG, KD-Tree\cite{ram2019revisiting}) start from one or more entry points. Even if these entry points are far from the query vector, the algorithm can rapidly converge to the neighborhood of the query within a small number of hops by traversing graph edges or tree branches, resulting in a sharp drop in query distance. The second phase is the fine-grained exploration phase for top-$k$ results. At this point, the algorithm has already reached the local neighborhood of the query, but must continue expanding multiple neighboring branches to ensure that no closer candidates are overlooked. This phase involves broader exploration with higher uncertainty in search paths, and may include nodes farther from the query vector, leading to distance fluctuations or even increases.

Although prior research has optimized the first phase using techniques such as vector compression and routing prediction\cite{wang2024starling,ge2013optimized,jegou2010product}, the second phase remains challenging due to frequent candidate expansion, scattered access patterns, and poor data locality. As a result, disk I/O becomes the dominant bottleneck to query performance. Therefore, this work focuses on optimizing the second phase by improving data locality and cache hit rates, thereby reducing disk accesses during queries and enhancing overall retrieval efficiency.

\section{Related Work}\label{sec:work}

\subsection{Vector Indexing}

With high-dimensional semantic vectors being widely used in tasks such as natural language processing, recommender systems, and graph neural networks\cite{openai2022gpt3embedding}, the design of efficient vector indexing structures to support large-scale, high-concurrency, and low-latency ANNS has become a core challenge for vector database systems. Traditional exact search methods suffer from the \textit{curse of dimensionality} in high-dimensional spaces, where both index construction and query efficiency deteriorate sharply, making them unsuitable for online service scenarios. To balance retrieval efficiency and accuracy, ANNS indexing techniques have emerged and gradually evolved into fundamental components of industrial-grade vector database systems such as Faiss\cite{douze2024faiss} and Milvus\cite{wang2021milvus}. Mainstream vector indexing approaches can be broadly categorized into three types.

\begin{itemize}
    \item \textbf{Hashing-based indices} (e.g., LSH\cite{jafari2021survey}). Multiple hash functions map vectors into different buckets, and candidate search is performed only within the same bucket. This method has low computational overhead but requires a large number of hash tables to achieve high recall, leading to increased memory consumption and system maintenance costs, which limits scalability.
    \item \textbf{Quantization-based indices} (e.g., IVF\_PQ\cite{jegou2010product}, SCANN\cite{hassantabar2021scann}). The vector space is compressed or partitioned to reduce computation and storage costs. These methods are widely deployed in industrial systems, such as Faiss, which accelerates candidate clustering using product quantization. However, in high-accuracy search tasks, quantization errors may cause similar vectors to be mapped to different cells, reducing recall and overall retrieval effectiveness.
    \item \textbf{Graph-based indices} (e.g., HNSW, NSG, DiskANN). A sparse, navigable neighbor graph is constructed, and heuristic traversal strategies are employed to enable efficient approximate search in high-dimensional spaces. Owing to their superior accuracy and scalability, graph-based methods have recently become the mainstream solution for large-scale vector retrieval tasks.
\end{itemize}

\subsection{Optimization Methods for Graph-Based ANNS}

Graph index methods have attracted extensive attention for their high accuracy and scalability. The core idea is to precompute partial neighborhood relationships during index construction, represent vectors as nodes, and connect similar vectors with edges to form a sparse directed graph. During query processing, the search starts from an entry node and progressively locates target neighbors through a neighbor-expansion algorithm. Among typical graph index methods, HNSW achieves efficient in-memory search through a multi-layer hierarchical structure, NSG constrains graph sparsity to optimize search path length and reduce redundant edges, and DiskANN proposes a disk-resident graph index that employs BeamSearch to support large-scale vector retrieval and has been integrated into industrial systems such as Milvus.

\stitle{Storage Optimization Techniques} Storage layout is a key design factor in disk-based graph index systems, as its physical organization directly determines I/O efficiency during retrieval. Current mainstream vector database systems mainly adopt two basic storage strategies, insertion-order storage and hash-distributed storage. Insertion-order storage organizes data linearly according to insertion time. While simple and offering high write throughput, it completely ignores vector similarity, causing physically adjacent locations to correspond to highly dissimilar vectors in high-dimensional space. This leads to scattered placement of neighbors and a large number of random I/O operations during queries. Hash-distributed storage distributes vectors uniformly using hash functions, effectively avoiding hotspot issues. However, it also disrupts the physical locality of adjacent nodes in the graph structure, intensifying cross-page accesses.

To address these issues, researchers have proposed various innovative solutions. For example, Facebook’s Faiss-IVFOPQ\cite{ge2013optimized} clusters similar vectors via inverted indexing (IVF) and compresses them using product quantization (PQ), thereby reducing both the number of disk seeks and the amount of data transferred per I/O. The LENS system proposed by UC Berkeley mines access hotspots from query logs and proactively preloads them into memory. Experiments show that this approach improves cache hit rates by approximately 25\%\cite{cheng2023take}.

\stitle{Caching Optimization Techniques}
To reduce disk access latency during queries, modern ANNS systems generally incorporate caching mechanisms to mitigate frequent I/O caused by the uncertainty of graph traversal. Caching primarily works by preloading critical node data along the query path, thereby improving cache hit rates and query responsiveness. DiskANN, for instance, reduces disk accesses by preloading frequently accessed entry nodes and several of their multi-hop neighbors into memory. The Starling system introduces the concept of an in-memory navigation graph, which, under memory constraints, randomly samples a subset of nodes from the dataset and builds a lightweight navigation graph in memory using construction algorithms such as Vamana. This navigation graph provides queries with closer entry points, significantly shortening search paths on the disk-resident graph and effectively reducing I/O overhead.
\section{Overview of \gv}\label{sec:overview}

\begin{figure}[t]
    \centering
    \includegraphics[width=\linewidth]{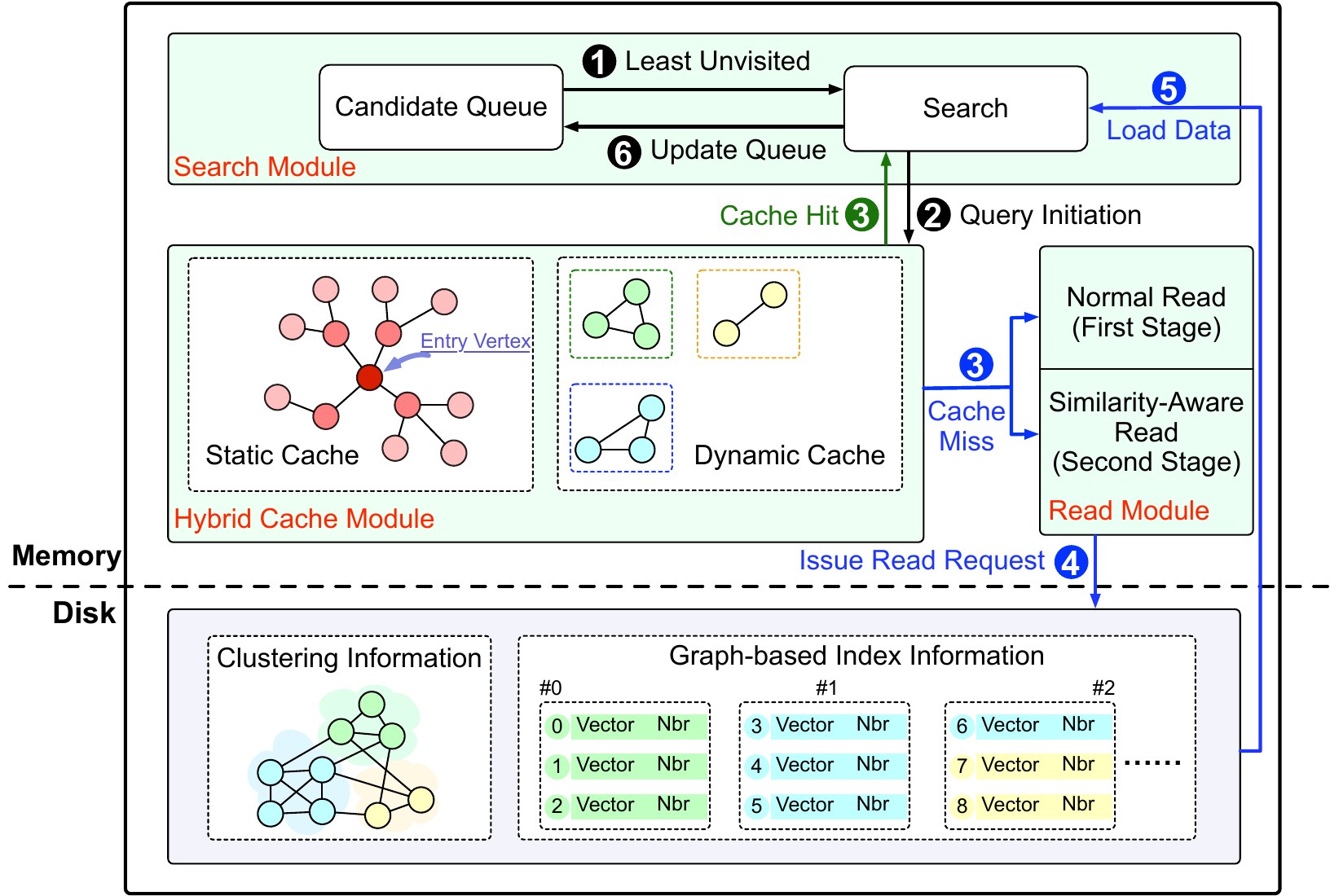} 
    \caption{ System architecture of \gv.}\label{fig:archi}
\end{figure}

This paper proposes \gv, an I/O-efficient caching strategy for vector neighbor search based on vector similarity. The overall framework of \gv is shown in Fig. \ref{fig:archi}. At the memory level, \gv designs a hybrid caching mechanism that combines static and dynamic components to adapt to different access behaviors during ANNS queries. In the static cache, the system preloads the entry point and several of its multi-hop neighbors according to a preset capacity, enabling rapid navigation to the candidate region near the query vector. In the dynamic cache, the system adaptively retains the pages containing nodes accessed during the query, as well as their adjacent pages on disk, to support the expansion of top-$k$ similar vectors. The detailed design of this mechanism will be presented in Section \ref{sec:hybrid}. At the disk level, \gv clusters the vectors in the neighborhood region of the query based on similarity, and then reorders the vectors in the original graph index accordingly. The reordered vectors are subsequently partitioned and stored on disk to enhance data locality and reduce cross-page accesses. This design will be elaborated in Section \ref{sec:reorder}.

Fig. \ref{fig:archi} illustrates the main steps of the search process. \ding{192} Select the closest unvisited node from the candidate queue for expansion. \ding{193} Issue a data read request to the hybrid cache module using the ID of the expansion node. \ding{194} The cache module checks whether the node resides in memory; if it is a hit, the node’s vector and adjacency information are returned, otherwise a disk access request is forwarded to the read module. \ding{195} The read module determines the disk loading strategy according to the current search phase. If the search is in the first phase, the disk page where the extension vertex is located is directly read. Otherwise, it employs a similarity-aware loading mechanism (see Section \ref{subsec:hybrid}) to load multiple pages containing the expansion node and its similar vectors. \ding{196} The retrieved vectors and adjacency information are loaded into memory. If the current phase is the second phase, the relevant pages are also written into the dynamic cache. \ding{197} Compute the exact distance between the expansion node and the query vector, and insert its neighbors into the candidate queue according to their PQ distances.

\section{Static-Dynamic Hybrid Caching Strategy}\label{sec:hybrid}

\subsection{Limitations of Existing Caching Strategies}\label{subsec:limit}

\begin{figure}[t]
    \centering
    \begin{subfigure}{.49\linewidth}
        \centering
        \includegraphics[width=\linewidth]{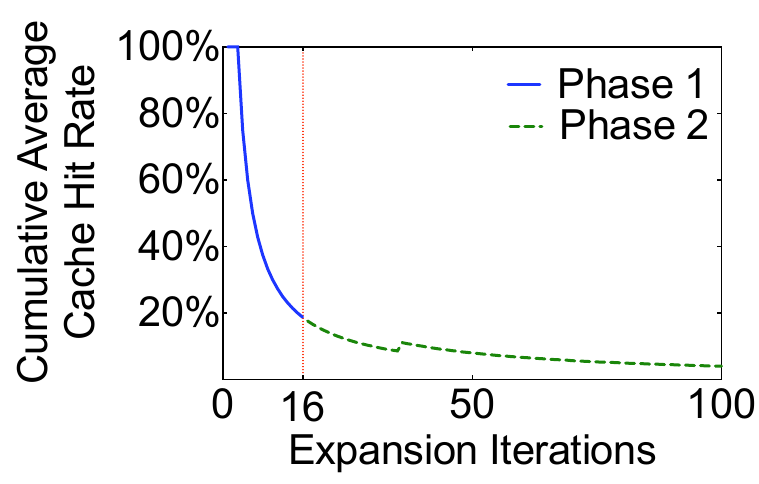}
        \caption{SIFT}\label{fig:limit-sift}
    \end{subfigure}
    \begin{subfigure}{.49\linewidth}
        \centering
        \includegraphics[width=\linewidth]{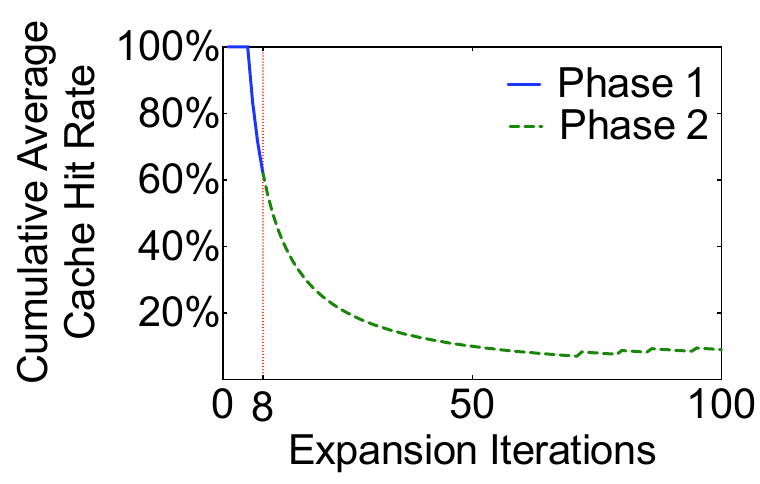}
        \caption{GIST}\label{fig:limit-gist}
    \end{subfigure}
    \caption{Static cache hit rate variation in DiskANN over 100 expansion steps ($k=10$).}
    \label{fig:limit-all}
\end{figure}

Traditional beam search algorithms typically adopt a static caching mechanism. During system initialization, the cache is preloaded with the entry vertex and several of its multi-hop neighbors based on the preset cache capacity, and the cached contents remain unchanged throughout the entire search process. This mechanism achieves a relatively high cache hit rate in the early phase of search because each query begins from the entry vertex, and the vertices in the first few hops are frequently accessed, thereby improving hit rates effectively. However, static caching cannot adapt to the actual expansion behavior of the query path, so its effectiveness is often limited to the initial few hops. As the search deepens, the hit rate drops rapidly, leading to a significant decline in memory access efficiency in the second phase.

We conducted experiments on DiskANN using the SIFT and GIST datasets, and the results are shown in Fig. \ref{fig:limit-all}. As the number of expansion steps increases, the static cache hit rate exhibits a clear power-law decay. Following the two-phase search division standard in Section \ref{subsec:two=phase}, we define the iteration at which the exact nearest neighbor is found as the turning point between the two phases, which is also marked in Fig. \ref{fig:limit-all}. It can be observed that when the search enters the second phase, the static cache hit rate decreases significantly compared to the first phase. In the first phase, the static cache hit rates are 19\% and 63\% for the two datasets, respectively; in the second phase, these rates drop sharply to only 4\% and 9\%. Furthermore, experiments show that the search time in the second phase often accounts for more than 80\% of the total query time, indicating that this phase has become the primary performance bottleneck limiting overall retrieval efficiency.

To gain deeper insights into the expansion characteristics of the second phase, we further analyze the distance distribution between expanded nodes and the query vector. As shown in Fig. \ref{fig:pre-stage}, the experimental results reveal that in this phase, the expanded nodes $p^\ast$ lie within a narrow distance range from the query vector $q$, exhibiting limited fluctuation and strong spatial concentration. Let $d(p^\ast, q)$ denote the distance between an expanded node and the query vector. According to the results in Fig. \ref{fig:pre-stage}, we have $d_{\min} < d(p^\ast, q) < d_{\max}$, where $d_{\min}$ and $d_{\max}$ represent the minimum and maximum distances observed during this phase. This implies that the query process in the second phase is effectively confined to an annular region centered at $q$ with a radius range of $[d_{\min}, d_{\max}]$. Compared to the entire vector space, the vectors within this local region are closer in distance and more likely to be accessed for distance computation and candidate expansion. Fig. \ref{fig:six-a} illustrates the structure of this annular region.
Building on this observation, if the query process in the second phase can focus on vectors within this region, reorganize them in order, and apply a corresponding physical storage layout, cache replacement operations would be confined to the disk pages covering this region. As a result, vectors loaded by each disk I/O would have a higher probability of being accessed, thereby significantly improving cache hit rates and reducing redundant I/O costs. However, existing disk-based graph index methods (e.g., DiskANN) still adopt a point-wise loading strategy during search, only the disk page containing the current expanded node is fetched, and once its vector and neighbor information are retrieved, the page is immediately evicted from memory. This strategy ignores the spatial clustering of expanded nodes in the second phase and fails to effectively leverage spatial locality along the search path, often leading to repeated disk I/O operations. Therefore, while the query process in the first phase can still rely on the traditional static caching mechanism, the second phase requires an enhanced strategy that exploits the locality of the annular region, enabling more efficient support for frequently accessed hot regions along the query path.

\subsection{Design of Query-Aware Hybrid Caching}\label{subsec:hybrid}

\begin{figure}[t]
    \centering
    \begin{subfigure}{\linewidth}
        \centering
        \includegraphics[width=.6\linewidth]{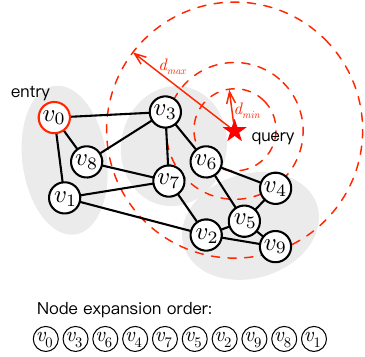}
        \caption{Range of vector expansion in the second-phase search}\label{fig:six-a}
    \end{subfigure}
    
    \vspace{1em}
    
    \begin{subfigure}{\linewidth}
        \centering
        \includegraphics[width=.9\linewidth]{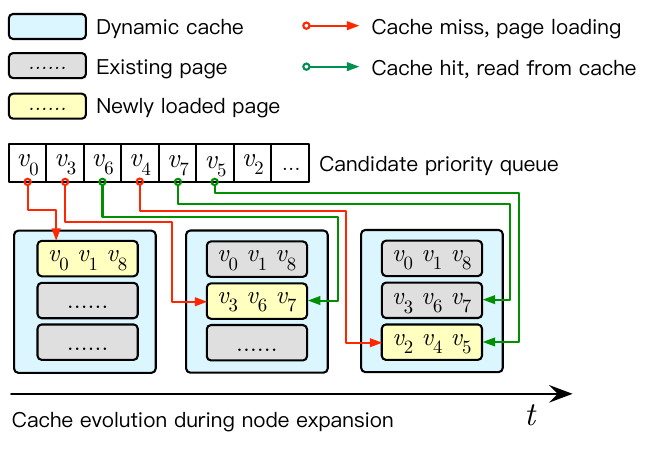}
        \caption{Vertex expansion process and dynamic cache updates}\label{fig:six-b}
    \end{subfigure}
    \caption{Illustration of dynamic caching.}
    \label{fig:six}
\end{figure}

Based on the analysis in Section \ref{subsec:limit}, we propose a hybrid caching mechanism that combines static and dynamic strategies. Specifically, search in the first phase adopts a static caching policy (consistent with traditional Beam Search) to accelerate the frequent accesses along the initial search path. Once the search enters the second phase, the system switches to a dynamic caching mechanism that focuses on optimizing locality-sensitive accesses. In particular, during the second phase, for a cache miss on an expanded vertex, \gv leverages the vertex’s position in the vector space to trigger a batch read operation, loading multiple neighboring vectors from disk into the dynamic cache. This strategy exploits spatial similarity among expanded vertices, increasing the likelihood that subsequent expansions hit the cache, thereby significantly reducing the performance overhead of frequent random I/O. As shown in Fig. \ref{fig:six-b}, when expanding vertex $v_3$, a cache miss occurs. The system then batches and loads the cache pages containing $v_3$, $v_6$, and $v_7$ into the dynamic cache based on their spatial proximity. Since vertices in the candidate priority queue have a high probability of being expanded, the previously loaded pages can be effectively reused. For example, when expanding $v_6$ and $v_7$, their pages have already been cached during the expansion of $v_3$, resulting in direct hits and avoiding redundant disk accesses, thus substantially improving overall query efficiency.

To support this dynamic caching mechanism, \gv introduces a similarity-aware reading strategy. To enable sequential batch loading, \gv optimizes the spatial layout of vector data during index construction, clustering and storing similar vectors together to ensure strong locality at the physical storage level (detailed in Section \ref{sec:reorder}). Consequently, during read operations, the system can locate related pages through sequential access, which incurs significantly lower overhead than traditional random I/O. The key to this similarity-aware strategy lies in dynamically determining the class and intra-class position of each expanded vertex, and computing the most suitable sequential read interval based on the current query context, thereby achieving locality-enhanced batch loading. Concretely, when expanding a target vertex, the system first identifies its class (the target class), then calculates the optimal sequential read interval based on the class size and the vertex’s relative position within it. This adaptive data reading mechanism follows three principles. \ding{192} Center the loading around the target vertex, prioritizing its adjacent vectors within the same class. \ding{193} If the target class cannot meet the demand, supplement from the target class and its neighboring classes. \ding{194} Ensure the read interval does not exceed boundary limits.

\begin{figure*}[t]
    \centering
    \includegraphics[width=.8\linewidth]{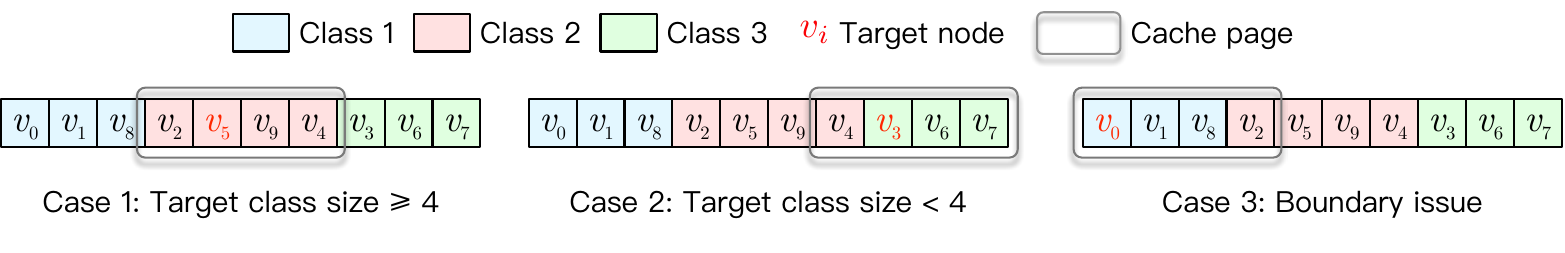} 
    \caption{Illustration of the similarity-aware reading mechanism.}\label{fig:seven}
\end{figure*}

Fig \ref{fig:seven} illustrates three typical scenarios commonly encountered in real-world applications, assuming a cache page size of four. In Case 1, the size of the target class is no smaller than the cache page capacity. Thus, centered on $v_{5}$, the system can directly load adjacent vertices $v_2$, $v_{5}$, $v_9$, and $v_4$ from the same class, completing an efficient cache fill. In Case 2, since the target class (Class 3) is too small to fill a cache page on its own, the system will focus on class 2 and also load the vertices ($v_4$) adjacent to the target vertex in Class 2. In Case 3, directly applying the above principle would cause the left boundary of the read interval to exceed the index file range. To avoid such abnormal behavior, the system performs boundary checking and adjustment to ensure correctness of the read operation.

Building on the static caching mechanism, \gv introduces the first query-aware dynamic cache. However, as the vector search proceeds, the data volume in the dynamic cache grows with the search path. Once it reaches the predefined capacity limit, cache replacement must be performed. Inspired by virtual memory management in operating systems, we implement three replacement strategies: First-In-First-Out (FIFO), Random, and Least Frequently Used (LFU). Based on the experimental results and analysis in Section \ref{subsec:expr-replace}, we select LFU as the default replacement strategy for dynamic caching.

Since we adopt different caching policies for the two search phases, accurately identifying the transition point between them is crucial for efficient search. We follow the method in PANNS\cite{yin2025panns}, which defines the transition as the moment when all top-$k$ candidates in the queue have been visited. However, this approach suffers from latency in detection, as the transition is often identified several iterations after it has actually occurred. As shown in Fig. \ref{fig:pre-stage}, on the SIFT and GIST datasets, the true transition points occur at the 16th and 8th iterations respectively (i.e., when the nearest neighbor is first reached), whereas PANNS detects them at the 27th and 16th iterations.

To overcome this lag, \gv introduces a tunable parameter $\theta \, (0 < \theta < 1)$. Specifically, we determine the transition point when all top-$\theta \cdot k$ candidates in the queue have been visited. The value of $\theta$ is estimated as follows. We randomly sample 1\% of queries from the dataset, record the actual transition round $k$ (i.e., the first iteration that reaches a top-$k$ neighbor), and compute the estimated transition round $k'$ using the PANNS method. For each query, we then calculate $\theta = k/k'$, and use these values to dynamically identify the transition point during search. This approach allows earlier detection of phase transitions, alleviating the lag issue significantly. In the experiments on Fig. \ref{fig:pre-stage}, \gv identifies transition points 9 rounds earlier on SIFT and 5 rounds earlier on GIST compared to PANNS, validating the effectiveness of this method.

Finally, the performance improvement of the dynamic caching mechanism not only depends on the caching strategy design but is also closely tied to the locality of the underlying storage layout. If similar vectors are clustered within the same disk page, sequential batch reads are more likely to load vectors that will soon be accessed, thereby improving cache hit rates. Hence, optimizing the storage layout to enhance I/O locality becomes a key complementary problem beyond caching. The next section focuses on this issue, presenting our vector similarity-based index graph reordering method.

\section{Vector-Similarity-Based Reordering\eat{ of Graph Index}}\label{sec:reorder}

\subsection{Analysis of I/O Efficiency Issues}

In large-scale vector search tasks, disk I/O cost often becomes the key bottleneck limiting system performance, partly due to the low utilization efficiency of each individual I/O. Since disk access is performed at the granularity of pages, when retrieving the features or neighbors of a node, the system must load the entire page containing that node. If only a small number of vectors in that page actually participate in the query, the rest of the data are redundant, leading to wasted bandwidth. To continue searching other relevant vectors, the system has to repeatedly load new disk pages, resulting in additional I/O overhead. Prior studies\cite{wang2024starling} show that in DiskANN, about 94\% of the vectors in each loaded page remain unused, and over 92.5\% of the query time is spent on disk I/O. This problem is particularly pronounced in the second phase of ANNS queries. Although a few nearest candidates are already identified at this phase, the system still needs to expand into the neighboring vector space to refine the results. If the expanded node and its neighboring vectors were colocated within the same disk page, a single I/O could fetch multiple potentially useful vectors, improving per-I/O efficiency and reducing redundant disk access. However, mainstream data layout strategies still largely organize pages according to graph connectivity. For example, the Starling system employs the Block Neighbor Frequency (BNF) algorithm to group graph-adjacent nodes together. Such methods ignore the structural discrepancy between graph topology and vector space, failing to guarantee the physical locality of similar vectors, and thus limiting improvements in I/O locality.

Fig. \ref{fig:eight} illustrates different storage layouts and their cache hit behaviors (assuming no static cache). Fig. \ref{fig:eight-a} shows the index graph structure; Fig. \ref{fig:eight-b} presents the storage layouts of DiskANN, Starling, and \gv (see Section \ref{subsec:reorder}), with each page assumed to hold at most three nodes; Fig. \ref{fig:eight-c} simulates the expansion process starting from the entry point and records the expansion nodes, the candidate priority queue, and whether the current page hits the next expansion node. During the experiment, starting from entry node $v_0$, the system expands $v_0$, $v_3$, and $v_6$ in sequence, maintaining the candidate queue. Since $v_6$ is closest to the query, the search enters the second phase after expanding $v_6$. At this point, the candidate queue becomes $\{v_4, v_7, v_5, v_8, v_1\}$, with $v_4$ as the next expansion node. Under DiskANN’s layout, because $v_6$ and $v_4$ are stored on different pages, two separate I/Os are needed. Similarly, Starling’s layout also places them on different pages, again requiring two I/Os. By the end of the search, both layouts suffer from low per-I/O utilization due to suboptimal partitioning.
Although increasing cache capacity or adopting asynchronous prefetching can alleviate I/O overhead, a more effective and lower-cost solution lies in designing better index layouts. Such layouts can significantly boost performance by improving locality, without relying on expensive runtime optimizations.

\subsection{Reordering Method of Graph Index}\label{subsec:reorder}

\begin{figure*}[t]
    \centering
    \begin{subfigure}{.28\linewidth}
        \centering
        \includegraphics[width=\linewidth]{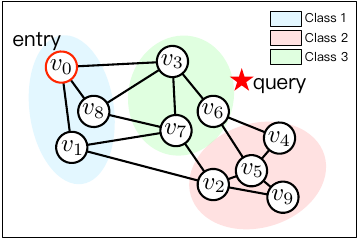}
        \caption{Index graph structure}\label{fig:eight-a}
    \end{subfigure}
    \hfill
    \begin{subfigure}{.28\linewidth}
        \centering
        \includegraphics[width=\linewidth]{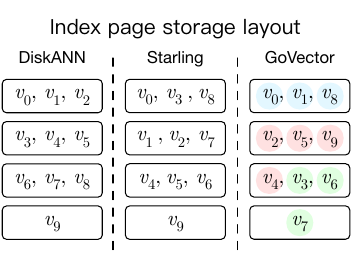}
        \caption{Storage layouts of different methods}\label{fig:eight-b}
    \end{subfigure}
    \hfill
    \begin{subfigure}{.4\linewidth}
        \centering
        \includegraphics[width=\linewidth]{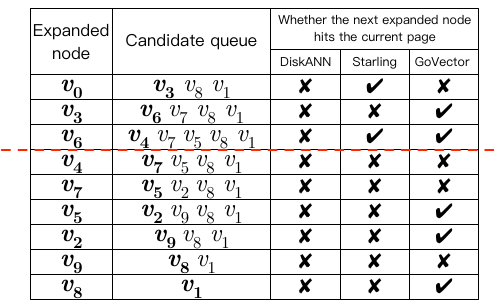}
        \caption{Cache hit behaviors under different methods}\label{fig:eight-c}
    \end{subfigure}
    \caption{Similarity-aware reordering of index graphs.}
    \label{fig:eight}
\end{figure*}

To overcome the limitations of existing layout methods, we propose \gv, an index graph layout optimization strategy that integrates vector similarity with storage reordering to improve disk access efficiency and overall retrieval performance. Specifically, \gv performs index graph reordering in two stages.

\begin{itemize}
    \item \textbf{Similarity clustering stage.} Based on Euclidean distance in the vector space, all vectors are partitioned into multiple high-similarity clusters using the $k$-means\cite{ahmed2020k} clustering algorithm. Within each cluster, vectors are spatially close to one another.
    \item \textbf{Locality optimization stage.} Building on the clustering results, and considering the index graph’s topology, vectors within the same cluster are placed on the same or physically adjacent disk pages whenever possible, thereby minimizing cross-page accesses during queries.
\end{itemize}

Fig. \ref{fig:eight} illustrates the reordering process and its effects. First, vectors are partitioned into clusters (class 1-3) using $k$-means in the original vector space, with the resulting index graph shown in Fig. \ref{fig:eight-a}. Within each cluster, vectors are further reordered according to graph connectivity. For example, in class 3 with vertices $\{v_{2}, v_{4}, v_{5}, v_{9}\}$, disk page size limits prevent all four from being stored together. Thus, $\{v_{2}, v_{5}, v_{9}\}$ are grouped on one page, since they are closer to the cluster centroid, while the peripheral vertex $v_{4}$ is placed on the next page. The final reordered layout is shown in Fig. \ref{fig:eight-b}.
During query processing, \gv effectively improves the utilization of each I/O operation. For instance, under a traditional beam search, when expanding vertex $v_{3}$, the system loads the page containing $v_{3}$. The next expansion, $v_{6}$, resides on the same page as $v_{3}$ in the \gv layout, thus avoiding an additional I/O and maximizing the previous bandwidth usage. Fig. \ref{fig:eight-c} compares cache hit behaviors under different layouts. DiskANN stores vectors either in insertion order or randomly, disregarding similarity, which disperses similar vectors across pages and incurs excessive I/O overhead. On the other hand, Starling places graph neighbors together on a page, reducing cross-page accesses to some extent. However, in the second phase of ANNS queries, expansions are similarity-driven rather than topology-driven. Since graph structure and similarity structure are not always aligned, Starling cannot accurately capture the spatial locality of the actual access path. In contrast, \gv first clusters vectors to identify high-similarity regions in the vector space, then performs local reordering and page assignment within each cluster by leveraging graph topology. This makes it more likely that similar vectors are loaded together on the same page during queries. Such a layout better aligns with the similarity-driven expansions in the second phase, significantly increasing I/O utilization. As a result, the next expansion vertex is more likely to reside on the same page as the current one, thereby reducing I/O overhead and improving query performance.

\section{Experiment}\label{sec:expr}

\subsection{Experimental Setup}

\stitle{Environment} All experiments were conducted on a high-performance server equipped with an Intel® Xeon® Gold 6248R processor (3.00GHz, 48 cores), 32GB DDR4 memory (3200MT/s), and two 1.7TB SSDs with a maximum sequential read/write bandwidth of 500MB/s. The operating system was Ubuntu 22.04 LTS, and the compiler version was GCC 11.4.0.

\stitle{Datasets} We used six publicly available real-world vector datasets, as listed in Table \ref{table:dataset}. These datasets cover diverse modalities such as images, text, audio, and word embeddings, with vector dimensionality ranging from 128 to 960. They have been widely used in the evaluation of existing ANNS systems.

\begin{table*}[htbp]
\centering
\caption{Experimental Datasets} \label{table:dataset}
\begin{tabular}{lccccc}
\toprule
Dataset   & Type  & Dimensionality & \#Vectors   & \#Queries & Content \\
\midrule
SIFT\cite{jegou2009searching}      & float & 128            & 1,000,000   & 10,000    & Image \\
Text2Img\cite{simhadri2022results}  & float & 200            & 1,000,000   & 1,000     & Image-Text \\
DEEP\cite{yang2024fast}      & float & 256            & 1,000,000   & 1,000     & Image \\
Word2Vec\cite{yang2024fast}  & float & 300            & 1,000,000   & 1,000     & Word Embedding \\
MSONG\cite{yang2024fast}     & float & 420            & 994,185     & 1,000     & Audio \\
GIST\cite{jegou2009searching}      & float & 960            & 1,000,000   & 1,000     & Image \\
\bottomrule
\end{tabular}
\end{table*}

\stitle{Baselines and Parameter Settings}
We compared \gv with two representative disk-based ANNS systems, DiskANN and Starling.
\begin{itemize}
    \item \textbf{DiskANN} is a graph-based disk-resident ANNS method. It adopts a greedy search strategy that progressively approaches the query vector starting from a pre-selected entry node along graph adjacencies. To reduce I/O overhead, DiskANN uses a static cache to preload frequently accessed entry nodes and several of their neighbors into memory. In our experiments, we set the cache size to 1\% of the index file, the default query size to Top-100, the neighbor degree $R=32$, and the number of search threads $T=32$.
    \item \textbf{Starling} is a recently proposed disk-resident index graph system that optimizes both storage and query paths through segmented data layouts. It employs a block-level search strategy, loading data page by page to reduce path length and I/O frequency. Furthermore, Starling applies a reordering algorithm (BNF strategy) based on topological structure to cluster neighboring nodes within the same disk page, thereby improving I/O utilization. In our experiments, we used the default BNF reordering strategy, with other parameters set to match DiskANN.
    \item \textbf{\gv} is an efficient hybrid caching strategy. By combining static and dynamic caching, it improves cache hit rates and overall query throughput without additional memory overhead. Its core idea is to optimize the physical layout of the index based on vector similarity, clustering vectors that are likely to appear in the same query path into the same disk page, thereby enhancing locality (in contrast to Starling’s topology-based reordering). In our experiments, the static-to-dynamic cache ratio was set to 2:8, while other parameters were consistent with DiskANN.
\end{itemize}

\subsection{Overall System Performance}

\begin{figure*}[t]
    \centering
    \begin{subfigure}{.7\linewidth}
        \centering
        \includegraphics[width=\linewidth]{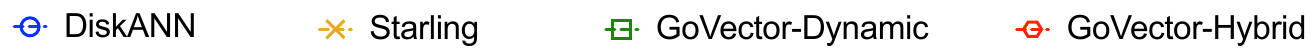}
    \end{subfigure}
    \\
    \vspace{0.2em}
    \begin{subfigure}{.26\linewidth}
        \centering
        \includegraphics[width=\linewidth]{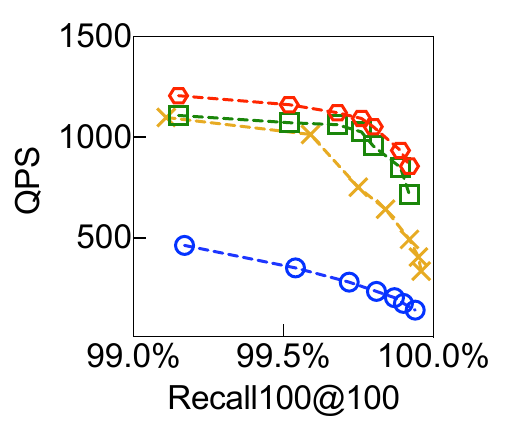}
        \caption{SIFT}
    \end{subfigure}
    \begin{subfigure}{.26\linewidth}
        \centering
        \includegraphics[width=\linewidth]{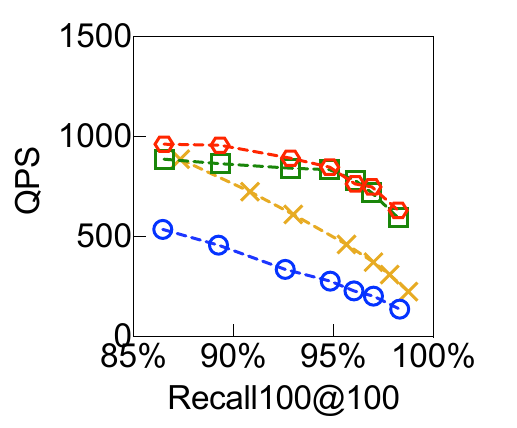}
        \caption{Text2Img}
    \end{subfigure}
    \begin{subfigure}{.26\linewidth}
        \centering
        \includegraphics[width=\linewidth]{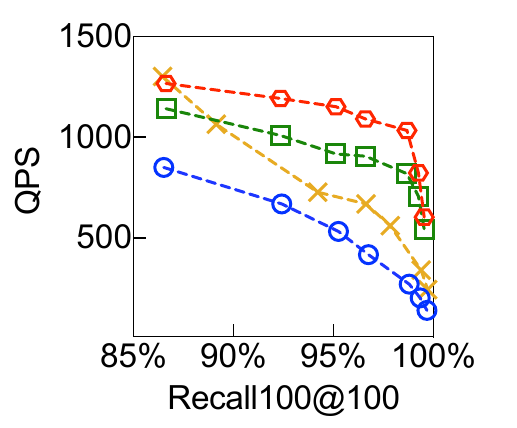}
        \caption{DEEP}
    \end{subfigure}
    \\
    \begin{subfigure}{.26\linewidth}
        \centering
        \includegraphics[width=\linewidth]{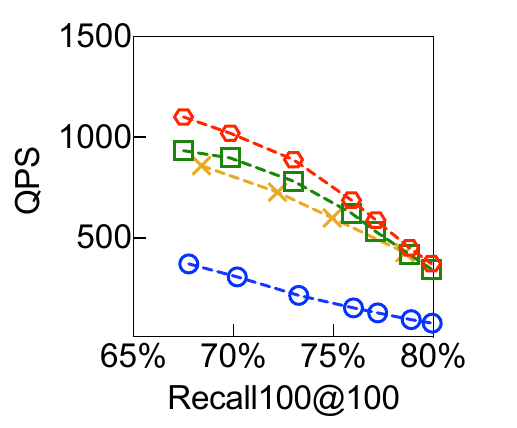}
        \caption{Word2Vec}
    \end{subfigure}
    \begin{subfigure}{.26\linewidth}
        \centering
        \includegraphics[width=\linewidth]{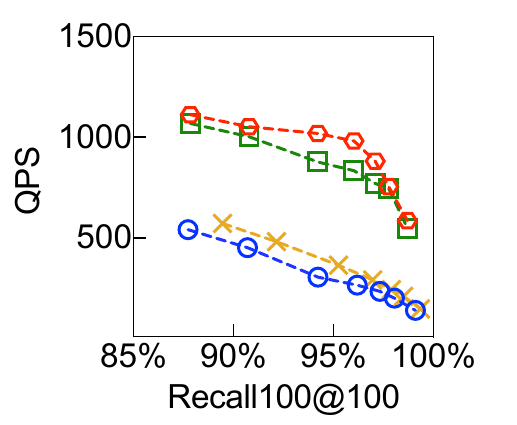}
        \caption{MSONG}
    \end{subfigure}
    \begin{subfigure}{.26\linewidth}
        \centering
        \includegraphics[width=\linewidth]{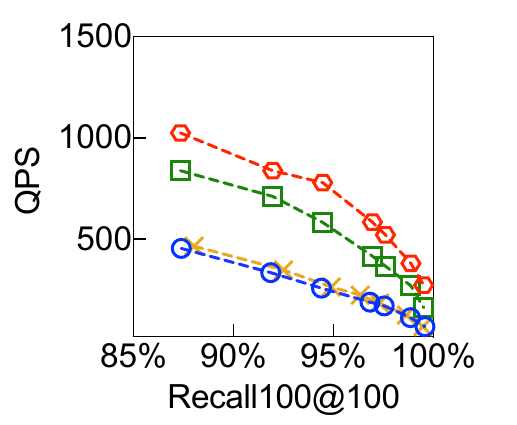}
        \caption{GIST}
    \end{subfigure}
    \caption{Performance comparison of \gv and other methods.}
    \label{fig:expr-overall}
\end{figure*}

Fig. \ref{fig:expr-overall} compares the performance of different ANNS methods in terms of Queries Per Second (QPS) and recall. \gv-Hybrid represents \gv with a combined static and dynamic caching strategy (with a 2:8 ratio of static to dynamic cache), while \gv-Dynamic represents a purely dynamic version of \gv without static caching (i.e., with zero statically cached nodes).

The experimental results show that \gv consistently achieves superior search performance compared to other ANNS methods, with its advantage becoming more pronounced under high-recall scenarios. Specifically, when recall is no less than 90\%, \gv-Hybrid achieves a QPS improvement of 2.61$\times$-4.59$\times$ over DiskANN, 1.10$\times$-3.97$\times$ over Starling, and 1.06$\times$-1.50$\times$ over \gv-Dynamic. Moreover, as data dimensionality increases, \gv maintains high search efficiency, whereas Starling’s performance remains close to that of DiskANN. This is largely attributed to GoVector’s efficient hybrid caching strategy and similarity-aware layout optimization.

It is worth noting that in a few low-recall scenarios (such as with the Text2Img and DEEP datasets), Starling achieves slightly higher QPS than \gv. This is because in these scenarios, the first phase of the search dominates the overall runtime, while \gv’s dynamic caching mechanism mainly optimizes the efficiency of the second phase, meaning its benefits are less pronounced. Additionally, we observe that in low-recall scenarios, \gv-Dynamic performs significantly worse than \gv-Hybrid. The main reason is that in the first phase, \gv-Hybrid leverages static caching to preload multi-hop neighbors of the entry nodes, thereby improving hit rates, reducing disk I/O accesses, and boosting overall performance. In contrast, under high-recall scenarios, as the search queue expands, the second phase becomes the main bottleneck. In this case, dynamic caching plays the dominant role, leading to similar performance between \gv-Dynamic and \gv-Hybrid.

\subsection{Analysis of Static-Dynamic Cache Ratios}

\begin{figure}[t]
    \centering
    \begin{subfigure}{.49\linewidth}
        \centering
        \includegraphics[width=\linewidth]{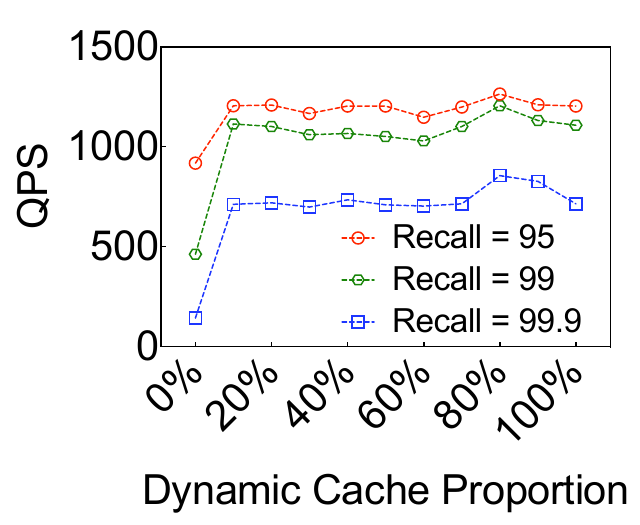}
        \caption{SIFT}
    \end{subfigure}
    \begin{subfigure}{.49\linewidth}
        \centering
        \includegraphics[width=\linewidth]{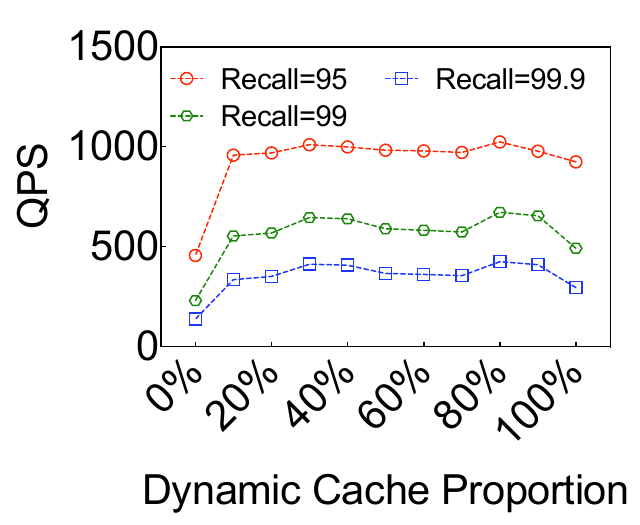}
        \caption{GIST}
    \end{subfigure}
    \caption{Search performance under different cache ratio configuration.}
    \label{fig:expr-ratio}
\end{figure}

We evaluated the impact of different static-to-dynamic cache ratios on search performance using two public datasets, SIFT and GIST, across various recall targets. The experimental results are presented in Fig. \ref{fig:expr-ratio}. Based on the data, we draw the following observations.

\textit{Benefits of rational cache allocation.} 
The results show that when the static-to-dynamic cache ratio is set to 2:8 (i.e., 80\% dynamic cache), the system achieves the best performance across all recall levels. In GoVector, static cache mainly accelerates the first-phase search, while dynamic cache optimizes data access in the second phase. When the dynamic cache share exceeds 80\%, the reduced capacity of the static cache prevents it from effectively storing entry points and their multi-hop high-frequency neighbors, making it difficult to hit relevant data early in the search. This increases the latency of the first phase, which dominates overall query throughput. Although dynamic cache still benefits the second phase, the slowdown in the first phase lowers the overall QPS. This indicates that an appropriate balance between static and dynamic cache yields an effective trade-off between performance and resource utilization.

\textit{Effectiveness of dynamic cache.} Even a small amount of dynamic cache can substantially improve performance under the same recall level. For example, on the SIFT dataset, increasing the dynamic cache share from 0\% to 10\% improves performance by 1.31$\times$ to 5.04$\times$. This demonstrates the efficiency of dynamic caching in capturing hot spots and adapting to evolving search paths.

\textit{Stability and robustness of dynamic cache.} Further analysis shows that the performance trends of dynamic cache remain consistent across recall levels, reflecting strong stability. Moreover, under any non-zero dynamic cache setting, the system outperforms the purely static configuration (i.e., 0\% dynamic cache). This highlights the robustness and adaptability of dynamic caching across a wide range of retrieval accuracy requirements.

\subsection{Impact of Different Cache Replacement Strategies}\label{subsec:expr-replace}

\begin{figure}[t]
    \centering
    \begin{subfigure}{.52\linewidth}
        \centering
        \includegraphics[width=\linewidth]{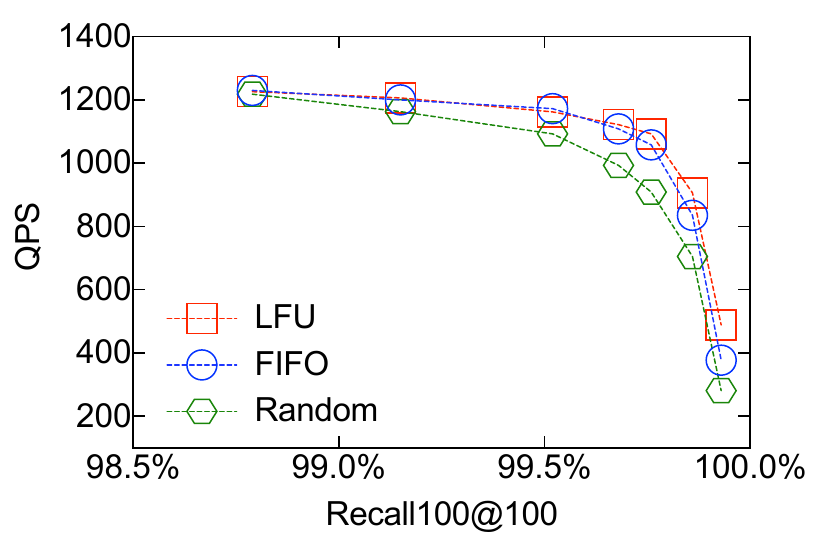}
        \caption{Search performance}\label{fig:expr-replace-a}
    \end{subfigure}
    \begin{subfigure}{.46\linewidth}
        \centering
        \includegraphics[width=\linewidth]{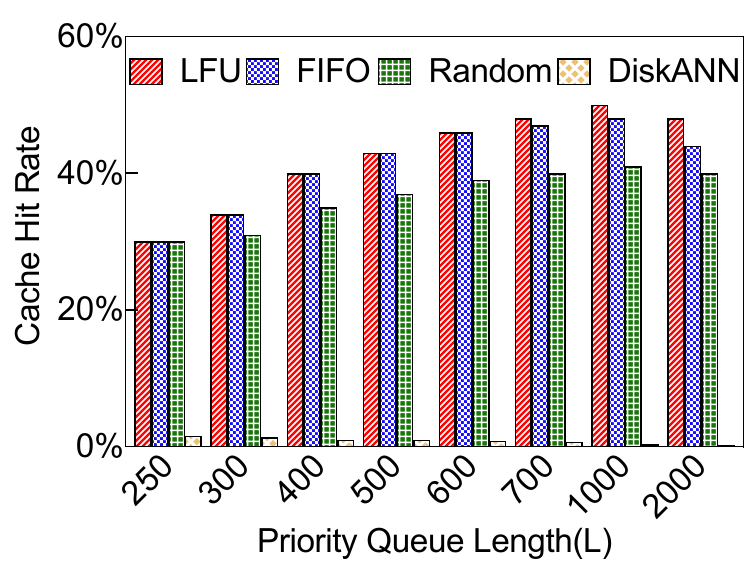}
        \caption{Cache hit rate}\label{fig:expr-replace-b}
    \end{subfigure}
    \caption{Performance of different cache replacement strategies.}
    \label{fig:expr-replace}
\end{figure}

To explore the optimal replacement strategy for dynamic caching, we evaluated three common policies on the SIFT dataset: Least Frequently Used (LFU), First-In-First-Out (FIFO), and Random. Fig. \ref{fig:expr-replace-a} reports their search performance under different recall targets. The results show that LFU achieves the best overall performance. Fig. \ref{fig:expr-replace-b} further presents the cache hit rates of these policies under varying priority queue lengths. First, as the queue length increases, GoVector’s dynamic cache consistently achieves significantly higher hit rates than the static cache employed by DiskANN. This demonstrates that dynamic caching is more effective in adapting to locality shifts during search, promptly capturing potential hot nodes, thereby reducing disk I/O and improving overall performance. In contrast, DiskANN’s static cache maintains a fixed set of nodes and cannot adapt to query-dependent access patterns, which limits both its hit rate and efficiency.

Further analysis shows that when the priority queue length is 250, all three replacement strategies achieve nearly identical hit rates and search performance. This is because no replacement operations are triggered at this phase, and the cache contents remain in their initialization phase. When the queue length increases to 300-600, LFU and FIFO perform similarly. The reason is that LFU assumes frequently accessed data will continue to be hot, whereas FIFO assumes recently loaded data are more likely to be reused. Under the current search workload, newly loaded nodes are indeed accessed more frequently, while earlier nodes are gradually evicted, making the two strategies converge in effect. However, when the queue length further grows to 700-2000, LFU begins to show clear advantages in both hit rate and search performance. This is because FIFO, relying solely on arrival time, tends to evict “historical hot” nodes that were loaded earlier but still accessed frequently, thereby lowering cache efficiency. LFU, on the other hand, preserves these frequently visited nodes more effectively, leading to higher hit rates and improved overall search performance.

\subsection{Impact of Different k}

\begin{figure}[t]
    \centering
    \begin{subfigure}{.49\linewidth}
        \centering
        \includegraphics[width=\linewidth]{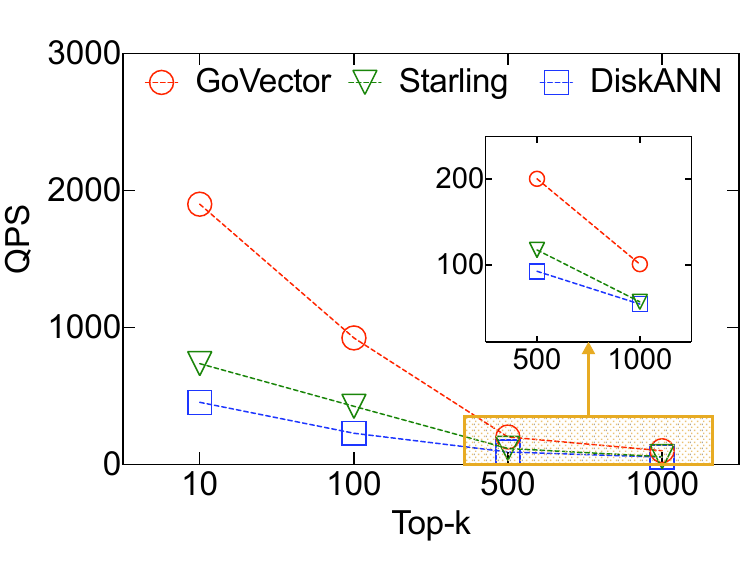}
        \caption{DEEP}
    \end{subfigure}
    \begin{subfigure}{.485\linewidth}
        \centering
        \includegraphics[width=\linewidth]{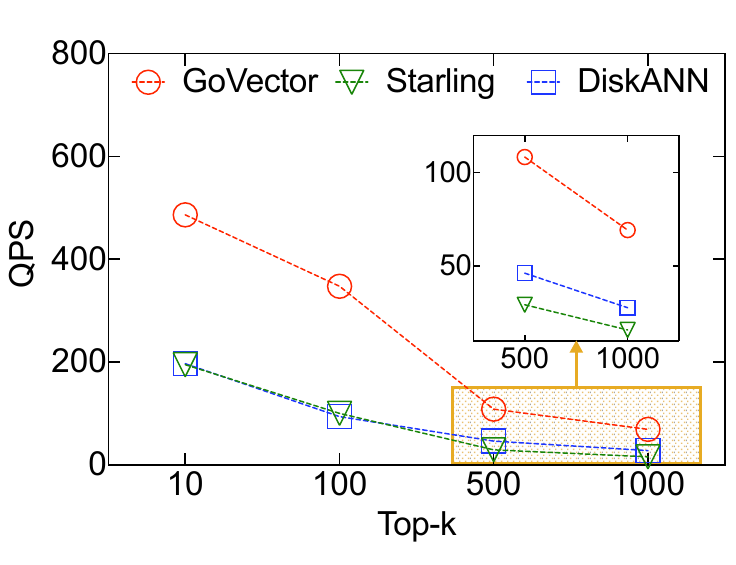}
        \caption{GIST}
    \end{subfigure}
    \caption{QPS performance of different $k$ values at 99\% recall.}
    \label{fig:expr-k}
\end{figure}

This section reports the query throughput (QPS) of GoVector (i.e., GoVector-Hybrid), Starling, and DiskANN under different Top-$k$ settings ($k$ = $10$, $100$, $500$, and $1000$), while ensuring a 99\% recall rate. The results are presented in Fig. \ref{fig:expr-k}. The results show that,
\begin{itemize}
    \item When $k = 10$, GoVector achieves 2.47$\times$-4.18$\times$ higher QPS than DiskANN and 2.49$\times$-2.58$\times$ higher than Starling.
    \item When $k = 100$, GoVector achieves 3.69$\times$-4.03$\times$ higher QPS than DiskANN and 2.18$\times$-3.46$\times$ higher than Starling.
    \item When $k = 500$, GoVector achieves 2.17$\times$-2.34$\times$ higher QPS than DiskANN and 1.71$\times$-3.69$\times$ higher than Starling.
    \item When $k = 1000$, GoVector achieves 1.86$\times$-2.50$\times$ higher QPS than DiskANN and 1.77$\times$-4.36$\times$ higher than Starling.
\end{itemize}

These results demonstrate that GoVector consistently delivers high and stable query performance across different candidate set sizes. It significantly improves system throughput while maintaining high recall, validating its generality and efficiency under diverse retrieval accuracy requirements.

\section{Conclusion}\label{sec:con}

This paper addresses the challenge of disk-based Approximate Nearest Neighbor Search (ANNS) and proposes GoVector, an I/O-efficient caching strategy guided by vector similarity. GoVector introduces a hybrid cache mechanism that integrates static and dynamic caching tailored to the two phases of ANNS search. The static cache preloads entry nodes and their multi-hop neighbors to accelerate the initial localization of candidate regions, while the dynamic cache adaptively stores candidate nodes and their vector-space neighbors during search to improve the hit rate in the similarity exploration phase.

In addition, GoVector incorporates a vector-similarity-aware graph reordering strategy and an adaptive data loading mechanism to enhance spatial locality of cached data, thereby further improving query throughput and I/O efficiency. Experimental results show that GoVector consistently outperforms state-of-the-art disk-based indexing methods (e.g., DiskANN and Starling) across multiple public datasets, demonstrating strong performance advantages.

One limitation of GoVector lies in the manual configuration of the static-to-dynamic cache ratio in the hybrid mechanism. Such fixed settings may fail to achieve optimal performance across diverse query workloads and data distributions. As future work, we plan to design query-aware and system-monitoring-driven adaptive cache adjustment strategies that automatically balance static and dynamic caches, thereby further improving the generality and robustness of GoVector.

\bibliographystyle{IEEEtran}  
\bibliography{ref} 

\end{document}